\documentclass[twocolumn]{aastex63}
\usepackage{amsmath}

\shorttitle{An alternative model for the orbital decay of M82 X-2}
\shortauthors{Chen}

\begin{document}


\title{An alternative model for the orbital decay
of M82 X-2: the anomalous magnetic
braking of a Bp star}

\author[0000-0002-0785-5349]{Wen-Cong Chen}
  \affil{School of Science, Qingdao University of Technology, Qingdao 266525, People's Republic of China; chenwc@pku.edu.cn}
  \affil{School of Physics and Electrical Information, Shangqiu Normal University, Shangqiu 476000, People's Republic of China}



\begin{abstract}
Recently, the first pulsating ultraluminous X-ray source M82 X-2 was reported to be experiencing a rapid orbital decay at a rate of $\dot{P}=-(5.69\pm0.24)\times 10^{-8}~\rm s\,s^{-1}$ based on seven years \emph{NuSTAR} data. To account for the observed orbital-period derivative, it requires a mass transfer rate of $\sim200\dot{M}_{\rm edd}$ ($\dot{M}_{\rm edd}$ is the Eddington accretion rate) from the donor star to the accreting neutron star. However, other potential models cannot be completely excluded. In this work, we propose an anomalous magnetic braking (AMB) model to interpret the detected orbital decay of M82 X-2. If the donor star is an Ap/Bp star with an anomalously strong magnetic field, the magnetic coupling between strong surface magnetic field and irradiation-driven wind from the surface of the donor star could cause an efficient angular-momentum loss, driving a rapid orbital decay observed in M82 X-2. The AMB mechanism of an Ap/Bp star with a mass of $5.0-15.0~M_{\odot}$ and a surface magnetic field of $3000-4500~\rm G$ could produce the observed $\dot{P}$ of M82 X-2. We also discuss the possibility of other alternative models including the companion star expansion and a surrounding circumbinary disk.
\end{abstract}

\keywords{X-ray binary stars (1811); Ultraluminous X-ray sources (2164); Orbital evolution (1178); Neutron stars (1108)}

\section{Introduction}
Ultraluminous X-ray sources (ULXs) are off-nucleus X-ray sources found in nearby galaxies,
and are generally thought to be accreting compact objects from a donor star \citep{fabb89}. It is well known that the accretion process of a compact object should be limited by the classical Eddington accretion rate ($\dot{M}_{\rm edd}$). However, isotropic X-ray luminosities of ULXs exceed the Eddington luminosity of stellar-mass black holes \citep[BHs,][]{fabb89}. Therefore, the accretors in ULXs were firstly proposed to be intermediate-mass BHs \citep[with a mass of $10^{2}-10^{5}~M_{\odot}$,][]{colb99,li04,kaar17}.
However, the discovery of the first pulsating ULX M82 X-2 implied that the accreting compact object is a neutron star (NS) \citep{bach14}. Subsequently, other six pulsating ULXs including NGC7793 P13 \citep{furs16,isra17a}, NGC5907 X-1 \citep{isra17b}, NGC300 X-1 \citep{carp18}, NGC 1313 X-2 \citep{sath19}, M51 ULX-7 \citep{rodr20}, and NGC7793 ULX-4 \citep{quin21} were discovered in succession. Furthermore, there exist three pulsating ULX candidates including M51 ULX-8 \citep{brig18}, SMC X-3 \citep{tsyg17}, and Swift J0243.6+6124 \citep{wils18}.

Study on the pulsating ULXs is very important in understanding astrophysical processes associated with strong gravitational field, stellar and binary evolution, accretion physics, magnetic field, and equation of state of NSs, hence these sources raised a keen interest of theoretical and observational astrophysicists. Binary population synthesis predicted that NS accretors dominate the ULX population in the solar metallicity environment \citep{shao15,wikt17}. However, all of NS ULXs appear as a geometrically beamed emission (less than 50\% of BH ULXs are beamed), resulting in an underrepresentation of NS ULX populations in observations \citep{wikt15,wikt19,misr24}. Based on the reported spin period, spin-up rate (the spin period derivative $\dot{P}_{\rm s}\approx-2\times 10^{-10}~\rm s\,s^{-1}$), and X-ray luminosity of M82 X-2 \citep{bach14}, some works proposed that the pulsar is a NS with a strong magnetic field of $\ga10^{13}~\rm G$ \citep{eksi15,dall15,mush15,kari16}, while other works inferred the NS to be a low-magnetic-field magnetar with a normal dipolar field and relatively strong multipole field \citep{tong15,chen17a}, a weak-field NS \citep{kluz15,king16}, or a NS with a traditional dipole field \citep{xu17}. Recently, \cite{king19} showed that observed X-ray luminosities, spin periods, and spin-up rates of  those pulsating ULXs are in agreement with the emission from NSs with normal magnetic fields ($\sim10^{11}-10^{13}~\rm G$). Subsequently, through 15 times observations by \emph{NuSTAR} throughout 2015 and 2016, \cite{bach20} found that M82 X-2 was experiencing a spin-down at an average rate ($\dot{\nu}\sim-6\times10^{-11}~\rm Hz\,s^{-1}$) between 2014 and 2016, providing a strong evidence that the pulsar is close to a spin equilibrium. Recently, \cite{liu24} reported that the spin-down
rate of M82 X-2 is $\dot{\nu}=-7.4\times10^{-11}~\rm Hz\,s^{-1}$ by measuring its pulsation back to 2005 and
2001 using Chandra and XMM-Newton archive data.

M82 X-2 was thought to be a NS high-mass X-ray binary (HMXB) including a donor star with a mass greater than $5.2~M_{\odot}$ in an orbit of $P=2.533$ days \citep{bach14}. Recently, \cite{bach21} measured the orbital-period derivative of M82 X-2 to be $\dot{P}/P=-8.20\pm0.34\times 10^{-6}~\rm yr^{-1}$ based on seven years data of \emph{NuSTAR}. This implies that the orbit of this source is shrinking at a rate of $\dot{P}=-(5.69\pm0.24)\times 10^{-8}~\rm s\,s^{-1}$. Then, they claimed that the observed orbital decay can be interpreted by a mass transfer from the more massive donor star to the less massive NS at a rate of $\sim200~\dot{M}_{\rm edd}$ ($\dot{M}_{\rm edd}=2.4\times10^{-8}~M_{\odot}\rm yr^{-1}$ is the Eddington accretion rate). Considering a slightly lower radiation efficiency or a massive outflow, such a mass transfer rate is sufficient to yield the observed high X-ray luminosity \citep[the peak isotropic X-ray luminosity of M82 X-2 is $1.8\times10^{40}~\rm erg\,s^{-1}$,][]{bach14} without beaming effect. However, \cite{king21} emphasized that a millennium of orbital-period changes is still too short to deduce the mass transfer rate, such a mass transfer rate deduced from a short-term observation is physically meaningless. If the observed orbital decay is a short-term phenomenon, Applegate's mechanism can be responsible for such a negative orbital-period derivative \citep{appl92}. However, the operation of this mechanism requires the donor star to have a strong magnetic field of $\sim1000~\rm G$ for the donor star. Actually, the magnetic coupling between the magnetic field and the stellar winds would cause a secular orbital decay rather than a short-term orbital shrinkage for the donor star with such a strong magnetic field (see also section 3).

If the orbital decay of M82 X-2 is a long-term phenomenon, other alternative models including the companion star expansion and a surrounding circumbinary disk can also work \citep{bach21}. Therefore, the real mechanism caused the orbital decay of M82 X-2 still remains controversial, and other alternative model cannot be completely ruled out. In this work, we propose an alternative scenario to account for the rapid orbital decay observed in M82 X-2. In Section 2, we discuss whether the mass transfer models including conservative and nonconservative models can work for M82 X-2. Section 3 describes the anomalous magnetic braking model and shows its predicted results. A brief discussion and conclusion are presented in Sections 4, and 5, respectively.

\section{Mass transfer model}
M82 X-2 was inferred to be a NS HMXB including a O/B giant donor star with a mass greater than $5.2~M_{\odot}$ \citep{bach14,bach21}. Using the detailed binary evolution model, some works perfectly reproduce the observed properties of this source \citep{frag15,quas19,misr20}. Therefore, we study its present evolution based on the NS HMXB scheme.

Analysis for the period modulation of M82 X-2 confirmed a near-circular orbit with an upper limit of 0.003 on eccentricity \citep{bach14}. Recently, the upper limit of the eccentricity was refined to be 0.0015 \citep{bach21}. For simplicity, we take a circular orbit for M82 X-2. The total orbital-angular momentum of a NS HMXB with a circular orbit can be expressed as
\begin{equation}
J= \mu a^{2}\frac{2\pi}{P}
\end{equation}
where $a$ and $P$ are the orbital separation and the orbital period of the binary, respectively; $\mu=M_{\rm ns}M_{\rm d}/(M_{\rm ns}+M_{\rm d})$ is the reduced mass of the binary ($M_{\rm ns}$ and $M_{\rm d}$ are the NS mass and the donor-star mass, respectively). Differentiating equation (1) and inserting Kepler's third law, the orbital-period derivative satisfies
\begin{equation}
\frac{\dot{P}}{P}=3\frac{\dot{J}}{J}+3(q\beta-1)\frac{\dot{M}_{\rm d}}{M_{\rm d}}+\frac{(1-\beta)\dot{M}_{\rm d}}{M_{\rm ns}+M_{\rm d}},
\end{equation}
where $\beta=-\dot{M}_{\rm ns}/\dot{M}_{\rm d}$ is the accreting efficiency of the NS, $q=M_{\rm d}/M_{\rm ns}$ is the mass ratio of the system, $\dot{M}_{\rm ns}$ and $\dot{M}_{\rm d}$ are the accretion rate of the NS and the mass-loss rate of the donor star, respectively. The mass-loss rate of the donor star consists of the mass-transfer rate ($\dot{M}_{\rm tr}$) and the wind-loss rate ($\dot{M}_{\rm w}$), that is $|\dot{M}_{\rm d}|=\dot{M}_{\rm tr}+\dot{M}_{\rm w}$.

Since $\dot{J}<0$, $\dot{M}_{\rm d}<0$, and $\beta\leq1$, the first and third terms on the right-hand side of equation (2) cause a negative orbital period derivative. The second term would contribute a positive (or negative) $\dot{P}$ if the term $q\beta<1$ (or $>1$). Therefore, the orbital evolution fate of a NS HMXB would depend on the competition between the first, second, and third terms in equation (2).

During the evolution of a NS HMXB, the orbital-angular-momentum-loss mechanisms includes gravitational radiation, mass loss, and other additional mechanisms. The orbital-period derivative of M82 X-2 due to gravitational radiation can be written as
\begin{equation}
\begin{aligned}
\dot{P}_{\rm gr}=-2.87\times10^{-14}\frac{(M_{\rm ns}/1.4~M_{\odot})(M_{\rm d}/10~M_{\odot})}{[(M_{\rm ns}/1.4~M_{\odot})+(M_{\rm d}/10~M_{\odot})]^{1/3}} \\
\left(\frac{2.533~\rm days}{P}\right)^{5/3} ~\rm s\,s^{-1}.
\end{aligned}
\end{equation}
Even if a large mass range of the donor star is considered, the orbital-period derivative is still 5-6 orders of magnitude smaller than the observed value. Therefore, we ignore the angular-momentum-loss rate due to gravitational radiation in the following discussion. However, the mass transfer plays a vital role in influencing orbital evolution of a HMXB. In this section, we consider two cases consisting of conservative and nonconservative mass transfer.

\subsection{conservative mass transfer}
In this case, $\dot{M}_{\rm ns}=-\dot{M}_{\rm d}$, i.e. $\beta=1$. In addition, conservative mass transfer implies that the orbital-angular-momentum is conservative, i.e. $\dot{J}=0$. Hence, equation (2) yields
\begin{equation}
\frac{\dot{P}}{P}=3\frac{\dot{M}_{\rm d}}{M_{\rm d}}(q-1).
\end{equation}
Because $\dot{M}_{\rm d}<0$, the orbit will shrink when the material is transferred from the more massive donor star to the less massive NS ($q>1$). Taking $M_{\rm ns}=1.4~M_{\odot}$, and $M_{\rm d}=10~M_{\odot}$, we plot the orbital-period derivative predicted by the conservative mass transfer model in Figure 1. We adopt an Eddington accretion rate of $\dot{M}_{\rm edd}=2.4\times10^{-8}~M_{\odot}\rm yr^{-1}$ same to \cite{bach21}. It is clear that the conservative mass transfer model with $\dot{M}_{\rm d}\sim180\dot{M}_{\rm edd}$ can produce the observed orbital-period derivative $\dot{P}=-(5.69\pm0.24)\times 10^{-8}~\rm s\,s^{-1}$ measured by \cite{bach21}. If the donor-star mass is smaller than $10~M_{\odot}$, it would require a high mass-transfer rate to produce the observed $\dot{P}$ according to equation (4).

Actually, the conservative mass transfer would experience dynamical instability when the material is transferred from the much massive donor star to the less massive NS \citep{hjel87,ivan04}. Even if the mass transfer could remain stable in a timescale of $10^{4}-10^{5}$ yrs, it still remains three puzzles. First, it is extremely challenging how to explain such a high accretion rate of $\sim180~\dot{M}_{\rm edd}$. Strong radiation pressure from the accreting NS should eject a fraction of the accretion material even if the Eddington accretion rate is ignored. Second, such a high accretion rate should produce a magnetospheric radius
\begin{equation}
\begin{aligned}
r_{\rm m}=3.8\times10^{6}\left(\frac{\dot{M}}{180~\dot{M}_{\rm edd}}\right)^{-2/7}\left(\frac{M_{\rm ns}}{1.4~M_{\odot}}\right)^{-1/7}\\
\left(\frac{BR^{3}}{10^{30}~\rm G\,cm^{3}}\right)^{4/7}~\rm cm,
\end{aligned}
 \end{equation}
where $B$ and $R$ are the magnetic field and the radius of the NS. However, the corotation radius of M82 X-2 is
$r_{\rm co}=2.1\times10^{8}(M_{\rm ns}/1.4~M_{\odot}^{1/3})~\rm cm$ \citep{chen17a}. The observations indicate that M82 X-2 is close to a spin equilibrium \citep{bach20,bach21}. According to $r_{\rm m}=r_{\rm co}$, the magnetic field of the NS can be estimated to be $B\approx1.1\times10^{15}~\rm G$, which is much higher than the estimated value by different models (see also the Introduction). Third, taking $\dot{M}_{\rm acc}=180~\dot{M}_{\rm edd}$, the total accretion luminosity of the accreting NS is given by $L_{\rm acc}=0.1\dot{M}_{\rm acc}c^{2}\approx2.4\times10^{40}~\rm erg\,s^{-1}$, which is greater than the peak isotropic X-ray luminosity of $L_{\rm iso}=1.8\times10^{40}~\rm erg\,s^{-1}$ \citep{bach14}. In fact, the mean X-ray luminosity ($\sim6.0\times10^{39}~\rm erg\,s^{-1}$) of M82 X-2 is slightly smaller than $1.8\times10^{40}~\rm erg\,s^{-1}$ \citep{brig16}. Therefore, it is highly unlikely that the mass transfer of M82 X-2 is conservative.

\subsection{non-conservative mass transfer}

According to the peak isotropic luminosity of M82 X-2 $L_{\rm max}=1.8\times 10^{40}~\rm erg\,s^{-1}$ \citep{kong07,feng10}, the mass transfer rate can be constrained to be $\dot{M_{\rm tr}}\simeq 36\dot{M}_{\rm edd}$ according to a super-Eddington accretion model \citep{king16,king17}. Actually, an analysis for Chandra's 15 yr data found that the X-ray luminosity of M82 X-2 varies from $\sim10^{38}$ to $\sim10^{40}~\rm erg\,s^{-1}$ \citep{brig16}. Therefore, such a  mass transfer rate of $\sim36\dot{M}_{\rm edd}$ estimated by the super-Eddington accretion model is just an upper limit.

Apart from a mass transfer, the massive donor star of the accreting NS in M82 X-2 should have a strong wind loss. \cite{taur00} found that isotropic winds from the vicinity of the NS in intermediate-mass X-ray binaries can help them to stabilize the mass transfer. A high NS mass can reduce the mass ratio of two components, and also help to avoid the unstable mass transfer \citep{shao12}. Recently, \cite{frag15} found that "fast winds" from the vicinity of the donor star can enhance the stability of the mass transfer because of small specific angular momentum.

We consider two angular-momentum-loss mechanisms including the "fast winds" of the donor star and isotropic winds ejected from the vicinity of the NS. The wind-loss rate of the donor star is $\alpha\dot{M}_{\rm d}$ ($\alpha$ is the wind-loss fraction of the donor star), and the loss rate of isotropic winds is $(1-\alpha-\beta)\dot{M}_{\rm d}$. Therefore, we have the total angular-momentum-loss rate as $\dot{J}=\dot{J}_{\rm dw}+\dot{J}_{\rm is}$, in which the angular-momentum-loss rate due to the donor-star winds is
\begin{equation}
\dot{J}_{\rm dw}=\alpha\dot{M}_{\rm d}\frac{M_{\rm ns}J}{(M_{\rm ns}+M_{\rm d})M_{\rm d}},
\end{equation}
and the angular-momentum-loss rate producing by the isotropic winds of the NS is
\begin{equation}
\dot{J}_{\rm is}=(1-\alpha-\beta)\dot{M}_{\rm d}\frac{M_{\rm d}J}{(M_{\rm ns}+M_{\rm d})M_{\rm ns}}.
\end{equation}
Inserting equations (6) and (7) into equation (2), we have
\begin{equation}
\frac{\dot{P}}{P}=\frac{\dot{M}_{\rm d}}{M_{\rm d}}\left[3(1-\alpha)(q-1)-(1-\beta)\frac{2q}{1+q}\right].
\end{equation}
When $\alpha=0$, and $\beta=1$, equation (8) reduces to equation (4), i.e. conservation mass transfer.

Since $\dot{M}_{\rm d}<0$, the first term $f(\alpha)=3(1-\alpha)(q-1)$ in the brackets of right-hand side of equation (8) produces a negative orbital-period derivative. However, the second term $f(\beta)=-2(1-\beta)q/(1+q)$ results in a positive orbital-period derivative. Therefore, a small $\alpha$ and a large $\beta$ tends to produce the negative orbital-period derivative with a large $|\dot{P}|$. However, the donor star of M82 X-2 is inferred to be a high (or intermediate) mass giant star, which should produce strong stellar winds, leading to a relatively large $\alpha$ (see also section 3).

Here, we consider two cases of non-conservative mass transfer model:
\begin{enumerate}
  \item Case 1: According to the super-Eddington accretion model, the maximum mass transfer rate was constrained to be $\dot{M_{\rm tr}}\simeq 36\dot{M}_{\rm edd}$ \citep{king16,king17}. If the accretion process of the NS is limited by the Eddington accretion rate, $\beta=-\dot{M}_{\rm ns}/\dot{M}_{\rm d}=-\dot{M}_{\rm edd}/\dot{M}_{\rm d}$. Since $|\dot{M}_{\rm d}|=\dot{M}_{\rm tr}+\dot{M}_{\rm w}$, the minimum wind-loss efficiency is $\alpha=1+36\dot{M}_{\rm edd}/\dot{M}_{\rm d}$.

  \item Case 2: Similar to \cite{bach21}, we ignore the upper limit of the mass-transfer rate predicted by the super-Eddington accretion model, and take $\beta=-\dot{M}_{\rm edd}/\dot{M}_{\rm d}$ and $\alpha=0$.
\end{enumerate}

According to equation (8), Figure 1 also illustrates the orbital-period derivatives predicted by two cases of non-conservative mass transfer model. It is worth noting that Case 1 can not account for the observed orbital-period derivative of M82 X-2. A high mass-loss rate from the donor star would cause a small orbital-period derivative, which originates from a large wind-loss fraction $\alpha$ (a high mass-loss rate naturally results in a large $\alpha$ because $\alpha=1+36\dot{M}_{\rm edd}/\dot{M}_{\rm d}$). However, a large wind-loss fraction would lead to a small $|\dot{P}|$ according to equation (8).
Meanwhile, a high donor-star mass results in a large $|\dot{P}|$ for a same $\dot{M}_{\rm d}$.

In Case 2, a mass-transfer rate of $\sim200\dot{M}_{\rm edd}$ and $\sim270\dot{M}_{\rm edd}$ (see also figure 1) can produce the observed orbital-period derivative when the donor-star mass is $M_{\rm d}=10~M_{\odot}$ and $5~M_{\odot}$, respectively. Because the mass-growth rate of the NS is limited by the Eddington accretion rate, a massive outflow occurs from the vicinity of the accreting NS. The $|\dot{P}|$ predicted by Case 2 is similar to the conservative case. Since $\beta<1$, the second term in equation (2) leads to a positive $\dot{P}$. To produce a same $\dot{P}$, it requires a more high mass-transfer rate to compensate the positive $\dot{P}$ resulting from the second term. Therefore, the mass-transfer rate of Case 2 is slightly higher than that of
the conservative case. Similar to the conservative mass transfer, it also requires an ultra-strong magnetic field of $\ga10^{15}~\rm G$ to produce a large magnetospheric radius that is compatible with the corotation radius, resulting in a
spin-equilibrium situation of the NS \citep{bach20,bach21}.

\begin{figure}
\centering
\includegraphics[width=1.15\linewidth,trim={0 0 0 0},clip]{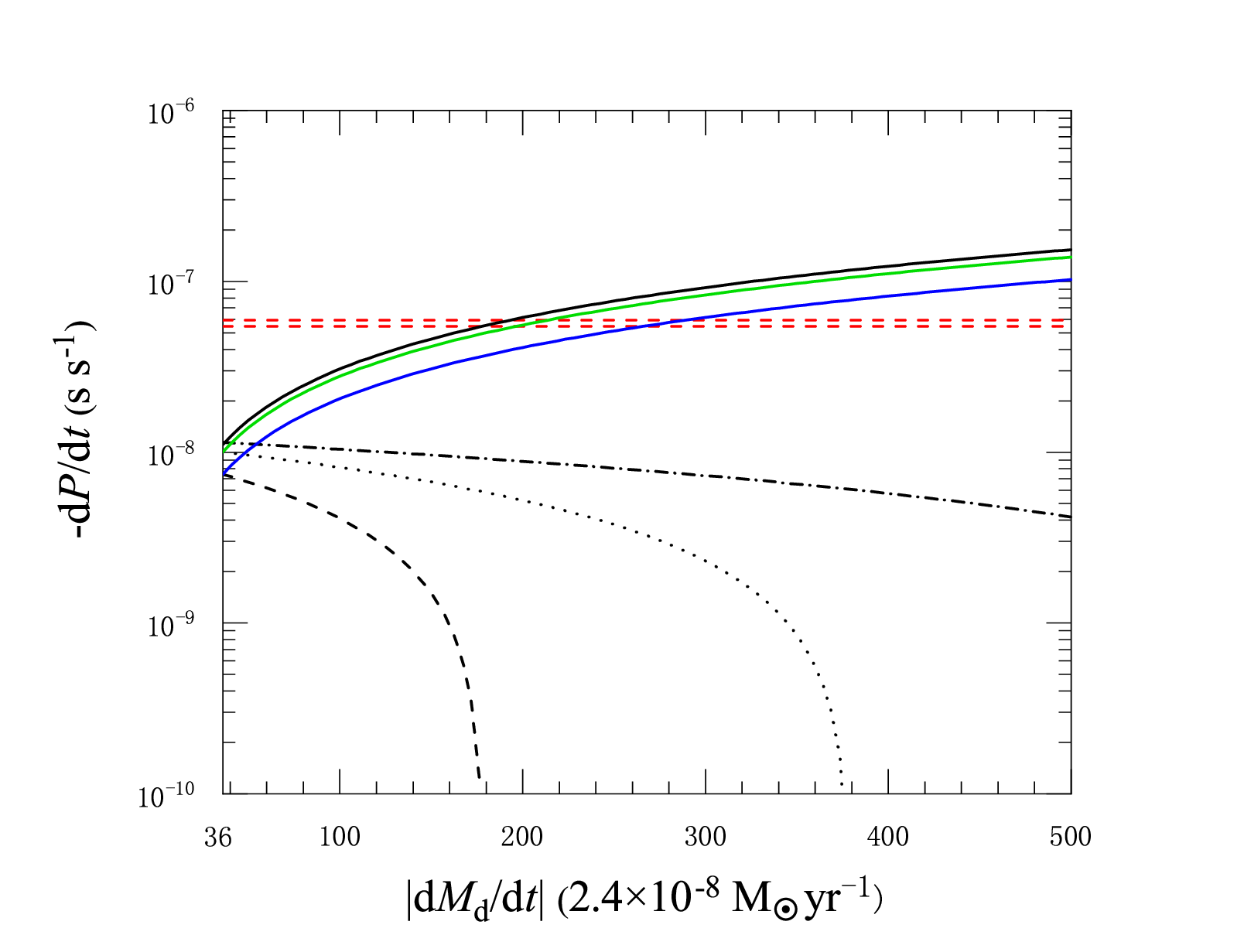}
\caption{Predicted orbital-period derivatives as a function of the mass-loss rate of the donor star. The black solid curve corresponds to the conservative mass transfer model when $M_{\rm d}=10~M_{\odot}$, i.e. $\alpha=0$, and $\beta=1$. The black dashed, dotted, and dashed-dotted curves represent the Case 1 of non-conservative mass transfer model when $M_{\rm d}=5, 10$, and $20~M_{\odot}$, respectively. The blue and green solid curves represent the Case 2 of non-conservative mass transfer model when $M_{\rm d}=5$ and $10~M_{\odot}$, respectively. Two horizontal red dashed lines represent the observed orbital-period derivative of M82 X-2.} \label{fig:orbmass}
\end{figure}

\begin{figure}
\centering
\includegraphics[width=1.15\linewidth,trim={0 0 0 0},clip]{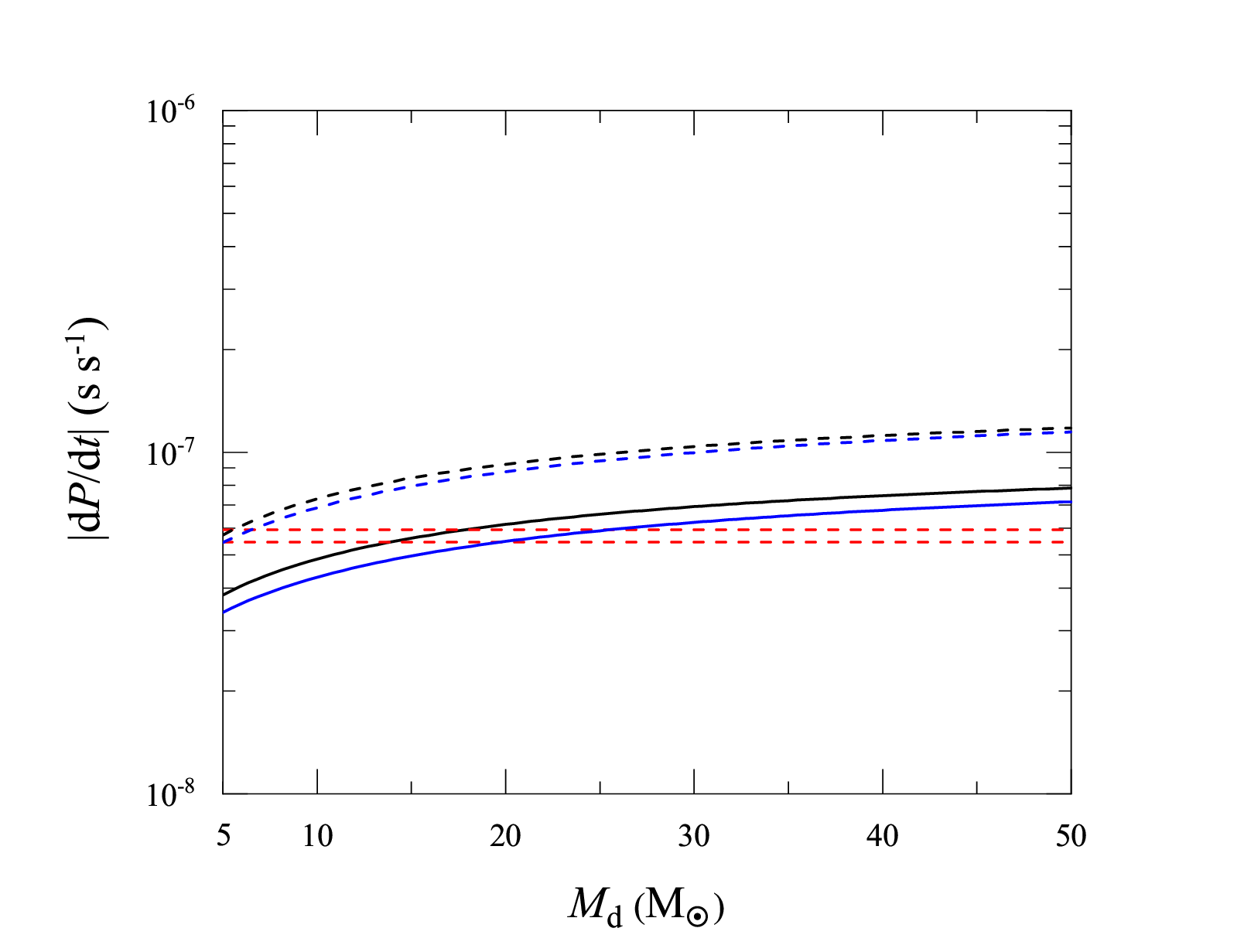}
\caption{Orbital-period derivatives producing by the AMB mechanism in the $\dot{P}-M_{\rm d}$ diagram for different surface magnetic fields of the donor star. The black solid and dashed curves denote the case with a surface magnetic field of $B_{\rm s}=3000$ and $4500~ \rm G$ when $M_{\rm ns}=1.4~M_{\odot}$, respectively; while the blue solid and dashed curves denote the case with a surface magnetic field of $B_{\rm s}=5000$ and $8000~ \rm G$ when $M_{\rm ns}=1.8~M_{\odot}$, respectively. The horizontal red dashed lines represent the observed orbital-period derivative of M82 X-2.} \label{fig:orbmass}
\end{figure}

\section{Anomalous magnetic braking model}
The magnetic braking (MB) mechanism can not operate for normal intermediate-mass stars because of the absence of
a convective envelope. However, a fraction of intermediate-mass stars, that is the so-called Ap and Bp stars, is found to possess anomalously strong surface magnetic fields of $10^{2}-10^{4}~\rm G$, which were proposed to be remnants of the star's formation, that is a 'fossil' magnetic field \citep{moss89,brai04}. In the A stars population, this small proportion is about $5\%$ \citep{land82,shor02}. In X-ray binaries, the magnetic coupling between strong surface magnetic field and X-ray irradiation-driven wind from the surface of the donor star could cause a plausible angular-momentum loss, that is the anomalous magnetic braking \citep[AMB,][]{just06}.

Assuming that a fraction $f$ of the X-ray luminosity irradiating on the donor star gets rid of the gravitational binding energy, and converts into the kinetic energy of the wind (with a velocity given by the escape
speed from the donor's surface), the irradiation-driven wind-loss rate obeys the following relation as \citep{chen16}
\begin{equation}
fL_{\rm acc}\frac{ R_{\rm d}^{2}}{4 a^{2}}=\frac{GM_{\rm d}\dot{M}_{\rm w}}{R_{\rm d}},
\end{equation}
where $f$ is a wind driving energy efficiency, $R_{\rm d}$ is the radius of the donor star, and $L_{\rm acc}$ is the intrinsic accretion luminosity of the ULX.

Actually, it is possible to produce a high wind-loss rate for the donor star in an ULX. On the giant branch, the wind-loss rate can be estimated by $\dot{M}_{\rm w}\sim4\times10^{-13}(L/L_{\odot})(R_{\rm d}/R_{\odot})(M_{\odot}/M_{\rm d})~M_{\odot}\,\rm yr^{-1}$, where $L$ is the luminosity of the donor star \citep{kudr78}. Assuming that $M_{\rm ns}=1.4~M_{\odot}$, and $M_{\rm d}=10~M_{\odot}$, the orbital separation is derived to be $a\approx17.6~R_{\odot}$. Therefore, the Roche-lobe radius of the donor star is $R_{\rm L}=a\times0.49q^{2/3}/[0.6q^{2/3}+{\rm ln}(1+q^{1/3})]\approx9.7~R_{\odot}$ \citep{eggl83}. Because the donor star fills its Roche lobe, $R_{\rm d}=R_{\rm L}$. For a star with a mass of $10~M_{\odot}$ on the giant branch, $L\sim10^{4}L_{\odot}$ \citep[see also figure 1 in][]{hurl00}, hence the wind-loss rate is $\dot{M}_{\rm w}\sim4\times10^{-9}~M_{\odot}\,\rm yr^{-1}$. If the irradiation-driven winds is included, the accretion luminosity of M82 X-2 is $L_{\rm acc}=bL_{\rm iso}$, here $b=0.06$ is the beaming factor \citep{king16}. Taking a maximum isotropic X-ray luminosity $L_{\rm iso}=1.8\times10^{40}~\rm erg\,s^{-1}$, it yields $L_{\rm acc}\sim10^{39}~\rm erg\,s^{-1}$, and derives the maximum wind-loss rate of the donor star producing by the irradiation process of the accreting NS to be $\sim6.0\times10^{-7}~M_{\odot}\,\rm yr^{-1}$ for a wind-driving efficiency $f=0.001$ according to equation (9). Such an irradiation-driven wind-loss rate is two orders of magnitude higher than the normal wind-loss rate of the donor star on the giant branch. Therefore, we ignore the normal wind loss in the AMB model.

During the wind outflows, the magnetic energy density providing by the Bp star
is greater than the kinetic energy density of winds inside the magnetospheric radius, and can tie up the charge particles in the magnetic field lines to co-rotate with the star. According to the balance between the magnetic energy density and
the kinetic energy density, the magnetospheric radius of the donor star is derived as \citep{just06}
\begin{equation}
r_{\rm m}=\frac{B_{\rm s}^{1/2}R_{\rm d}^{13/8}}{(GM_{\rm d})^{1/8}\dot{M}_{\rm w}^{1/4}},
\end{equation}
where $B_{\rm s}$ is the surface magnetic field of the donor star. In a NS HMXB, the magnetic coupling between strong surface magnetic field and irradiation-driven wind results in the donor star to spin down. However, the tidal interactions between the two components would continuously spin the donor star up to a synchronous rotation with the orbital motion \citep{patt84}. Such a spin-up of the donor star indirectly consumes the orbital angular momentum.

It is generally thought that the irradiation-driven winds are
ejected at the magnetospheric radius, the rate of angular-momentum loss induced by the AMB is given by
\begin{equation}
\dot{J}_{\rm mb}=-\dot{M}_{\rm w}r_{\rm m}^{2}\frac{2\pi}{P}
=- \frac{B_{\rm s}R_{\rm d}^{13/4}\dot{M}_{\rm w}^{1/2}}{(GM_{\rm d})^{1/4}}\frac{2\pi}{P}.
\end{equation}
Inserting $\dot{M}_{\rm w}$ derived from equation (9) into equation (11), we have
\begin{equation}
\dot{J}_{\rm mb}=-\frac{B_{\rm s}(fL_{\rm acc})^{1/2}R_{\rm d}^{19/4}}{2(GM_{\rm d})^{3/4}a}\frac{2\pi}{P}.
\end{equation}

To power an ULX, the donor star should fill its Roche lobe, hence its radius can be written as
\begin{equation}
R_{\rm d}=a f(q),
\end{equation}
where $f(q)=0.49q^{2/3}/[0.6q^{2/3}+{\rm ln}(1+q^{1/3})]$ \citep{eggl83}. From equations (2), (12), and (13), the orbital-period derivative due to the AMB is expressed as
\begin{equation}
\dot{P}_{\rm mb}=-\frac{3B_{\rm s}(fL_{\rm acc})^{1/2}f(q)^{19/4}}{2(2\pi)^{7/6}G^{1/6}}\frac{(M_{\rm ns}+M_{\rm d})^{19/12}}{M_{\rm ns}M_{\rm d}^{7/4}}P^{13/6}.
\end{equation}

The wind driving energy efficiency is $f\sim0.001$ \citep{just06}, which has a bit of uncertainty. An analysis for Chandra's 15 yr data found that the isotropic X-ray luminosity of M82 X-2 varies from $\sim10^{38}$ to $\sim10^{40}~\rm erg\,s^{-1}$ \citep{brig16}. Approximately, we take a rough mean X-ray luminosity $L_{\rm X}\sim10^{39}~\rm erg\,s^{-1}$, hence the accretion luminosity is $L_{\rm acc}=bL_{\rm X}\sim10^{38}~\rm erg\,s^{-1}$. In equation (14), $fL_{\rm acc}$ is a degenerate parameter, thus we set $fL_{\rm acc}=10^{35}~\rm erg\,s^{-1}$. In principle, there exist a bit of uncertainty in the factor $fL_{\rm acc}$. However, the orbital period derivative weakly depends on it because $\dot{P}\propto(fL_{\rm acc})^{1/2}$. Taking $M_{\rm ns}=1.4~M_{\odot}$ and $M_{\rm d}=10.0~M_{\odot}$, and $B_{\rm s}=3500~\rm G$, the orbital period derivative of M82 X-2 is $\dot{P}=-5.6\times10^{-8}~\rm s\,s^{-1}$, which is approximately consistent with the observed value $\dot{P}=-(5.69\pm0.24)\times 10^{-8}~\rm s\,s^{-1}$ \citep{bach21}.

Figure 2 displays the orbital-period derivatives producing by the AMB mechanism in the $\dot{P}-M_{\rm d}$ diagram.
For a same surface magnetic field, the $\dot{P}$ tends to increase when the donor-star mass increases. To produce the observed
orbital period derivative, a low-mass donor star requires a strong surface magnetic field. For example, the donor star with a mass of $5.0~M_{\odot}$ and $15.0~M_{\odot}$ requires a surface magnetic field of $B_{\rm s}=4500$ and $3000~\rm G$, respectively. The surface magnetic field should be in the range of $\sim3000-4500~\rm G$ if the donor star of M82 X-2 has a mass of
$5.0-15.0~M_{\odot}$. The NS mass also influence the orbital period derivative, a high-mass NS tends to require a strong magnetic field to produce a same $\dot{P}$ for a same donor-star mass. To produce the observed $\dot{P}$, the donor star with a mass of
$5.0-15.0~M_{\odot}$ require a magnetic field of $\sim5000-8000~\rm G$ when $M_{\rm ns}=1.8~M_{\odot}$. Therefore, the AMB mechanism could successfully account for the observed $\dot{P}$ of M82 X-2 for a plausible range of parameters including the surface magnetic field and mass of Bp star, and the degenerate parameter $fL_{\rm acc}$.
\section{Discussion}
There exist four potential models interpreting the orbital decay of HMXBs including the mass transfer \citep{bach21}, the companion expansion \citep{levi93,levi00}, the circumbinary disk \citep{taur06}, and the AMB (this work).
To account for the observed $\dot{P}$ of M82 X-2, the mass transfer model requires a mass transfer rate of $\sim200~\dot{M}_{\rm edd}$ \citep{bach21}. When the massive donor star in HMXBs evolves into the Hertzsprung Gap, the thermal timescale
mass transfer can achieve such a high mass-transfer rate \citep{king01}. However, such a high mass-transfer rate would lead to an accretion luminosity as $L_{\rm acc}=0.1\dot{M}_{\rm edd}c^{2}[1+{\rm ln}(\dot{M}_{\rm tr}/\dot{M}_{\rm edd})]=8.4\times10^{38}~\rm erg\,s^{-1}$ \citep{king16}. This high mass-transfer rate leads to a low beaming factor as $b=73(\dot{M}_{\rm edd}/\dot{M}_{\rm tr})\approx0.002$, and the isotropic X-ray luminosity is $L_{\rm iso}=L_{\rm acc}/b\approx4.2\times10^{41}~\rm erg\,s^{-1}$ \citep{king09}, which is much higher than the observed peak X-ray luminosity. Furthermore, such a high mass transfer rate would produce a magnetospheric radius much smaller than the corotation radius unless the NS possesses an ultra-strong magnetic field of $B\approx1.1\times10^{15}~\rm G$, otherwise it is difficult to be compatible with the observation that M82 X-2 is close to a spin equilibrium \citep{bach20,bach21}. Through the detection of cyclotron resonance scattering features, \cite{brig18} discovered that the neutron star in M51 ULX-8 may possess a surface magnetic field of $\sim10^{15}~\rm G$, providing a hint that neutron stars in ULXs may be magnetars.

The correlation between $\dot{P}/P$ and the X-ray luminosity could be a probe to discriminate the mass transfer and the AMB models. In the mass transfer model, $\dot{P}/P\propto\dot{M}_{\rm d}\propto L_{\rm X}$, while this correlation change into $\dot{P}/P\propto L_{\rm X}^{1/2}$ in the AMB model according to equation (14). Certainly, such a test requires a long-term data spanning decades years.

\subsection{Companion star expansion}
Five eclipsing HMXBs (LMC X-4, Cen X-3, 4U 1700-377, SMC X-1, and OAO 1657-415) were also detected a significant orbital
period decay with a $\dot{P}/P=(1.0-3.5)\times10^{-6}~\rm yr^{-1}$ \citep{fala15}. Some works proposed that the combined contribution of the growth of the moment of inertia (expansion of the donor star) and stellar wind loss of a donor star could
be responsible for the orbital decay in HMXBs \citep{levi93,bago96}. When the companion star evolves to a late stage, its moment of inertia increases due to a shell expansion, leading to a spin-down of the stellar rotation. The tidal torque between two components would transfer orbital angular momentum to the companion to spin up it back to synchronize the orbital motion, indirectly causing an orbital decay detected in LMC X-4 \citep{levi00}.

 In the companion-star expansion model, the orbital period derivative is governed by \citep{levi93}
\begin{equation}
\frac{\dot{P}}{P}\simeq-\frac{\dot{I}}{\mu a^{2}/3-I},
\end{equation}
where $I$ is the moment of inertia of the companion star. For a $15~M_{\odot}$ star, the increase rate of the companion star's
moment of inertia is about $\dot{I}/I\sim3\times10^{-7}~\rm yr^{-1}$ when the
center hydrogen is nearly exhausted \citep{levi00}. If $\mu a^{2}/(3I)\sim1$, it is possible to produce an orbital decay with $\dot{P}/P\sim10^{-6}~\rm yr^{-1}$ in HMXBs. It can distinguish the companion star expansion and the AMB models whether a NS HMXB with a early type companion star is detected a rapid orbital decay. However, this model cannot be applied in M82 X-2, in which the mass transfer is driven by the Roche lobe overflow. For a companion star filling its Roche lobe, its radius is equal to the Roche-lobe radius, which would shorten during the orbital decay. As a consequence, the moment of inertia of the companion star should reduce, not increase. Therefore, we emphasize that the companion star expansion model is invalid for M82 X-2, which is in contradiction with the discussion given by \cite{bach21}.

\subsection{Circumbinary disk}
\cite{bach21} discussed that the observed orbital decay of M82 X-2 can be caused by
an equatorial circumbinary (CB) disk launched by L2. The rate of angular momentum loss extracting by a surrounding CB disk
satisfies \citep{taur06}
\begin{equation}
\frac{\dot{J}_{\rm cb}}{J}=\frac{\delta\gamma(1+q)\dot{M}_{\rm d}}{M_{\rm d}},
\end{equation}
where $\delta$ is the fraction of the mass loss of the donor star that is fed into an equatorial CB disk with a radius of $a_{\rm r}=\gamma^{2} a$ \citep{heuv94}. Inserting equation (16) into equation (2), the orbital period derivative due to the existence of a CB disk obeys
\begin{equation}
\frac{\dot{P}}{P}=3\frac{\dot{J}_{\rm cb}}{J}=\frac{3\delta\gamma(1+q)\dot{M}_{\rm d}}{M_{\rm d}}.
\end{equation}
Taking $M_{\rm ns}=1.4~M_{\odot}$, $M_{\rm d}=10.0~M_{\odot}$, $\dot{M}_{\rm d}=180~\dot{M}_{\rm edd}$, and $\gamma=1.5$ \citep[similar to][]{sobe97}, it yields $\dot{P}/P=8.7\times10^{-6}~\rm yr^{-1}$ only if $\delta=0.55$, which is an ultra-high fraction of the transferred mass feeding into the CB disk.

The way testing the CB disk model is a direct detection in observations. The luminosity of the CB disk can be calculated from
its torque exerting on the binary, that is $L_{\rm cb}=2\pi T_{\rm cb}/P=2\pi \dot{J}_{\rm cb}/P$ \citep{spru01}. Therefore, the luminosity of the CB disk can be expressed as
\begin{equation}
L_{\rm cb}=\frac{\delta\gamma(1+q)\dot{M}_{\rm d}}{M_{\rm d}}\frac{4\pi^{2}\mu a^{2}}{P^{2}}.
\end{equation}
Adopting those same parameters like the previous paragraph, we have $L_{\rm cb}=2.7\times10^{35}~\rm erg\,s^{-1}$. Both the surface density and shear rate decrease with radius, hence the luminosity of the CB disk is mainly radiated at a zone near its inner edge \citep{spru01}. For M82 X-2, the effective emitting area is of the order $A\sim\pi a_{\rm r}^{2}=\pi\gamma^{2}a^{2}=1.0\times10^{25}~\rm cm^{2}$.
It can derive the effective temperature to be $T_{\rm eff}\sim4600~\rm K$ from Stefan-Boltzmann law, suggesting that CB
disk may be detectable in the optical band. We use the effective temperature and Planck's law to calculate the intrinsic monochromatic flux of the CB disk in different bands, then derive the detected flux and the visual magnitudes based on the effective emitting area and the distance. Adopting the the AB magnitude system \citep{fuku96} and a distance of $3.6~\rm Mpc$ \citep{kaar06}, without extinction the visual magnitudes of the CB disk are estimated to be $27.2,26.5$, and $26.1$ in $gri$ bands, respectively. Therefore, it is very challenging to confirm the existence of a CB disc around M82 X-2 due to its long distance.

\section{Conclusion}
In this work, we propose an alternative mechanism to account for the orbital decay of M82 X-2. If the donor star is a Bp star with an anomalously strong magnetic field, the magnetic coupling between strong surface magnetic field and irradiation-driven wind from the surface of the donor star could cause an efficient angular-momentum loss, driving a rapid orbital decay observed in M82 X-2. Taking $M_{\rm ns}=1.4~M_{\odot}$ and $fL_{\rm acc}=10^{35}~\rm erg\,s^{-1}$, the AMB mechanism of a Bp star with a mass of $5.0-15.0~M_{\odot}$ and a surface magnetic field of $3000-4500~\rm G$ could produce the observed $\dot{P}$ of M82 X-2. A low-mass donor star or a high mass NS tends to require a strong surface magnetic field.

Only if the NS possesses an ultra-strong magnetic field of $10^{15}~\rm G$, the mass transfer model can work for M82 X-2.
When $55\%$ of the mass loss of the donor star feeds into the disk, the CB disk model can also interpret the orbital decay of this source. However, the companion star expansion model is invalid for M82 X-2.

\acknowledgments {We thank the referee for a very careful reading and
constructive comments that have led to the improvement of the manuscript.
This work was partly supported by the National Natural Science Foundation
of China (under grant Nos. 12273014) and the Natural Science Foundation
(under grant No. ZR2021MA013) of Shandong Province.}

\end{document}